\documentclass[prd,preprint,showpacs,aps]{revtex4-1}
\usepackage[]{fontenc}
\usepackage[latin9]{inputenc}
\usepackage{amsmath}
\usepackage{graphicx}
\usepackage{amssymb}

\makeatletter
 
 \@ifundefined{textcolor}{}
 {%
   \definecolor{BLACK}{gray}{0}
   \definecolor{WHITE}{gray}{1}
   \definecolor{RED}{rgb}{1,0,0}
   \definecolor{GREEN}{rgb}{0,1,0}
   \definecolor{BLUE}{rgb}{0,0,1}
   \definecolor{CYAN}{cmyk}{1,0,0,0}
   \definecolor{MAGENTA}{cmyk}{0,1,0,0}
   \definecolor{YELLOW}{cmyk}{0,0,1,0}
 }

\usepackage{pxfonts}
\usepackage{hyperref}

\makeatother

\begin{document}

\pacs{04.62.+v, 04.70.Dy}

\title{Rotating Unruh-DeWitt Detector in Minkowski Vacuum}

\author{Tadas K Nakamura}

\affiliation{CFAAS, Fukui Prefectural University\\
 4-1-1, Kenjojima, Eiheiji, Fukui 910-1195, JAPAN}

\email{tadas@fpu.ac.jp}
\begin{abstract}
Response of a circularly rotating Unrh-DeWitt detector to the Minkowski
vacuum is investigated. What the detector observes depends on the
surface (three volume) to define it by the Hamiltonian. Detectors
in the past literature were defined on a surface of a constant Minkowski
time, and this is the reason why rotating detectors investigated so
far resister particles. No particle is detected by a detector defined
by the Hamiltonian on a surface normal to the detector's orbit, in
agreement with the global analysis of vacua. A detector with drift
motion superposed on the linear acceleration is also examined, to
find the same effect. 
\end{abstract}
\maketitle

\section{Introduction}

Thermalization of a vacuum due to acceleration has been first suggested
by Fulling \cite{Fulling1973}. Unruh \cite{Unruh1976} and DeWitt
\cite{dewitt1979general} have shown that particles with Planckian
distribution are actually observed by an accelerated detector in an
inertial vacuum (Minkowski vacuum); such a detector is called Unruh-DeWitt
detector. After that, numerous papers has been published on this topic
to this date (see \cite{takagi1986vacuum,Fulling2005,Crispino2008}
and references therein). Compared to the well established effect of
linear acceleration, the particle detection due to acceleration of
rotational motion is still controversial.

Letaw and Pfautsch \cite{Letaw1981} investigated vacua correspond
to various Killing flows in a flat spacetime. They have categorized
Killing flows into six classes, and examined the vacuum in each class.
Their conclusion is that there are only two inequivalent vacua in
a flat spacetime, which they called Minkowski vacuum and Fulling vacuum.
The vacuum with the Killing flow of circular rotation is the Minkowski
vacuum, which means an observer in rotational motion does not see
particles in a Minkowski vacuum. However, analysis using Unruh-DeWitt
detector indicates that a rotating detector will resister particles
\cite{Letaw1981a,Kim1987} (see also \cite{Bell1983,Bell1987}).

Davies \textit{et al}.\ \cite{Davies1996} found that the existence
of a static limit is necessary for the particle detection in a rotating
orbit. They have shown that a detector is not excited in the Minkowski
vacuum confined in a boundary inside the light cylinder. From this
result they suggested the possibility of detector excitation by negative
energy. It was Korsbakken and Leinaas \cite{Korsbakken2004} (referred
as Paper 1 hereafter) who revealed the actual process of a rotating
detector to observe particles; they found the detector is excited
by the emission of negative energy quanta.

It was argued in Paper 1 that some wave modes can have negative generalized
energy when the flow has a static limit. Here let us use more specific
words {}``Killing energy'' for what termed {}``generalized energy''
in Paper 1; it is the energy-momentum along the Killing flow in interest.
Two specific Killing flows are investigated in detail in Paper 1:
one corresponds to the spatial circular rotation and the other is
the flow with drift motion superposed on linear acceleration. The
excitation of detector due to the absorption of negative Killing energy
was found in both cases.

\bigskip{}

The present paper is stimulated by the results in Paper 1 and explore
the response of accelerating detectors. Our point is that the detector's
response depends on the choice of surface (three volume) to define
it, and the detector excitation by negative Killing energy does not
occur when we choose the surface appropriately.

The Unruh-DeWitt detector is a hypothetical monopole interacting with
the field at one volumeless point. The orbit of the detector is externally
given, and its internal energy is the energy measured in the frame
comoving with the orbit; it becomes Killing energy when the orbit
is along the Killing flow. The detector's interaction is represented
by a small interaction term added to the Hamiltonian of the whole
system. In general, a Hamiltonian is defined by the integration of
energy-momentum tensor over a surface of a constant time. Therefore,
what a detector observes is the Killing energy integrated over the
surface of Hamiltonian.

The Hamiltonian used in Paper 1 is on a surface of a constant Minkowski
time, which is not normal to the detector's orbit. The measured Killing
energy becomes a combination of energy and momentum integrated over
the surface of constant Minkowski time. This can be negative for waves
with large negative momentum, and the detector is excited by the emission
of such waves.

In contrast, the Killing energy is always positive when integrated
over a surface normal to the Killing flow, just like the pressure
(three dimensional momentum flow) is always positive. Therefore, a
detector does not perceive negative Killing energy when we define
it on the normal surface.

These two results do not contradict; two detectors observes different
physical quantities which do not have to agree. A similar situation
takes place for the acceleration superposed on the drift motion.

\bigskip{}

In the present paper we first investigate why and how the negative
Killing energy modes can exit as a result of the surface choice. As
we will see, the negative Killing energy occurs when a wave with phase
phase speed slower than the detector crosses the surface oblique to
the Killing flow.

Though the wave phase speed is usually faster than the speed of light
for planar waves, it can be locally slower in the cylindrical modes
for the rotational vacuum as in our case. However, it is not intuitively
easy to understand the underlying mechanism with the cylindrical modes
expressed with Bessel functions. Therefore, we firstly mimic the slower
phase speed using hypothetical planar waves with imaginary mass in
Section II. The detector is moving with inertial motion there. This
is fake but not entirely unrealistic; we can clarify the mechanism
of negative energy mode with it.

Then we move on to the accelerating detectors with realistic models.
The response of a rotating detector to the Minkowski vacuum is examined
in Section III; we find the detector will not be excited with an appropriate
choice of surface. In Section IV another similar motion of detector,
motion with drift superposed in acceleration namely, is investigated.
Again a properly defined detector has no excitation due to the negative
Killing energy; it detects only the particles with Planckian distribution
of the ordinary Unruh effect with Doppler shift. Section V is for
brief concluding remarks.

\section{Negative Killing Energy Mode}

Here in this section we examine how the negative Killing energy modes
occur with a simple and somewhat unrealistic model. To begin with,
let us clarify the definition of Killing energy. Let $\zeta_{\mu}$
be the unit vector in the direction of a Killing flow and $T_{\nu}^{\mu}$
be the energy-momentum tensor. What we call Killing energy in the
present paper is the energy-momentum along the Killing flow; its flux
$j_{\nu}$ is defined as \begin{equation}
j_{\nu}=\zeta_{\mu}T_{\nu}^{\mu}\,.\end{equation}
 The gross Killing energy is obtained by integrating the above flux
over a specific surface.

In the following we examine the Killing energy of a real valued two
dimensional field with wave equation \begin{equation}
\phi_{,tt}-\phi_{,xx}-m^{2}\phi=0\,\label{eq:wave}\end{equation}
where we write $\partial\phi/\partial t=\phi_{,t}$, etc., in shorthand.
We use the unit system of $c=\hbar=1$ (speed of light = Planck constant
= unity) throughout the present paper.

Suppose a hypothetical monopole detector with internal degree of freedom
$\mu$ is coupled with the above scalar field $\phi$ by a small coupling
constant $c$; the detector is moving along a fixed trajectory $(t,x)=(T(\tau),X(\tau))$,
where $\tau$ is the detector's proper time. The total Hamiltonian
of the system is given as \begin{equation}
H(t)=\int_{t=\textnormal{const}}\left[h_{\phi}+c\left(\frac{dT}{d\tau}\right)^{-1}{\mu}(\tau)\phi(t,x)\delta(x-X)\right]dx+\left(\frac{dT}{d\tau}\right)^{-1}H_{\mu}(\tau(t))\label{detector}\end{equation}
 where $h_{\phi}=\frac{1}{2}(\phi_{,t}^{2}+\phi_{,x}^{2}+m^{2}\phi^{2})$
is the Hamiltonian density of the field and $H_{\mu}(\tau)$ is the
internal Hamiltonian of the detector, which is independent of the
proper time $\tau$. The time evolution of the detector's internal
dynamics is along its propertied $\tau$, which can be inversely written
as a function of $t$; $\tau$ and $X$ in the above expression should
be understood as $\tau=\tau(t)$ and $X=X(\tau(t))$. The factor $(dT/d\tau)^{-1}$
in the coupling term comes from the same reason, i.e., the interaction
takes place along the detector's proper time $\tau$, while the Hamiltonian
is defined along the Minkowski time $t$. It should be noted that
the Hamiltonian is defined by the integration over a Cauchy surface
of $t=\textnormal{constant}$ which depends on the choice of $(t,x)$
coordinates.

The total Hamiltonian $H(t)$ is time dependent since $X$ depends
on time $t$. This means $\partial H/\partial t\ne0$. On the other
hand, the Killing energy is a conserved quantity since the detector
moves along the Killing flow. It is conserved locally due to the Noether's
theorem, so its integration over any Cauchy surface is also conserved.
Thus we can write \begin{equation}
\frac{\partial}{\partial t}\int_{t=\textnormal{const}}dx[\zeta_{t}(t,x)h_{\phi}-\zeta_{x}(t,x)p_{\phi}]=\frac{\partial}{\partial\tau}H_{\mu}\,,\label{eq:conserve}\end{equation}
 where $p_{\phi}=\phi_{,t}\phi_{,x}$ is the momentum flux across
the surface of constant $t$. We neglected in the above expression
the interaction energy, which vanishes by long time average. The above
expression means what the detector measures is the Killing energy
integrated over a surface of constant $t$.

\bigskip{}

Now let us suppose the wave field is in the vacuum and the detector
is in the state with lowest energy $E_{0}$; note that the detector's
energy $E$, i.e., the eigenvalue of $H_{\mu}$, is the energy in
the detector's frame since the time evolution with $H_{\mu}$ is determined
by the proper time. The detector moves along the Killing flow, thus
$E$ is the Killing energy. The state vector of the total system is
decomposed as \begin{equation}
\left|E,\Psi\right\rangle =\left|E_{0}\right\rangle \left|0\right\rangle \,.\end{equation}

The transition by the coupling occurs only for $\left|0\right\rangle \rightarrow\left|1_{k}\right\rangle $
($1_{k}$ denotes the state with one particle in mode $k$) when the
coupling constant $c$ is small enough (see, e.g., \cite{DeWitt2003}):
\begin{align}
A(E,1;E_{0},0) & =ic\left\langle E,1_{k}\right|\int_{-\infty}^{\infty}dt\,\int dx\,\left(\frac{dT}{d\tau}\right)^{-1}\mu(\tau)\phi(t,x)\delta(x-X)\left|E_{0},0\right\rangle \nonumber \\
\, & =ic\left\langle E,1_{k}\right|\int_{-\infty}^{\infty}d\tau\,\mu(\tau)\phi(T(\tau),X(\tau))\left|E_{0},0\right\rangle \nonumber \\
\, & =ic\left\langle E\right|\mu(0)\left|E_{0}\right\rangle \int_{-\infty}^{\infty}d\tau\, e^{i\Delta E\tau}\left\langle 1_{k}\right|\phi(T(\tau),X(\tau))\left|0\right\rangle \,,\label{eq:defA-1}\end{align}
 In the past literature, the transition amplitude $|A|^{2}$ expressed
with the Wightman function is often used to reach the same conclusion.
However, we use the above expression because emission/absorption of
quanta is more transparent in this form; one can obtain the same result
with the Wightman function.

The field can be expanded as \begin{equation}
\phi=\left(a_{k}^{+}e^{-i\omega t}+a_{k}^{-}e^{i\omega t}\right)\, e^{ikx}\,,\end{equation}
 where $\omega>0$ and $a_{k}^{+}$ {[}$a_{k}^{-}${]} are the annihilation
{[}creation{]} operators for the particles in mode $k$. This definition
of annihilation and creation operators is based on the Hamiltonian
on the surface of constant $t$, therefore, the detector's excitation
calculated by these operators is in response to the Killing energy
integrated over that surface as in (\ref{eq:conserve}).

Since the terms with annihilation operator $a_{k}^{+}$ vanishes for
the vacuum state $|0\rangle$, the transition coefficient in (\ref{eq:defA-1})
reduces to \begin{equation}
\left\langle 1_{k}\right|\phi(T(\tau),X(\tau))\left|0\right\rangle =\exp i(\omega T(\tau)+kX(\tau))\,.\label{eq:creation}\end{equation}
 Suppose the detector is moving with a constant velocity $(\zeta_{0t},\zeta_{0x})$,
i.e., $(T,X)=(\zeta_{0t}\tau,\zeta_{0x}\tau)$. Then we have \begin{equation}
\int_{-\infty}^{\infty}d\tau\, e^{i\Delta E\tau}\left\langle 1_{k}\right|\phi(T(\tau),X(\tau))\left|0\right\rangle =\delta(\Delta E+\zeta_{0t}\omega+\zeta_{0x}k)\label{transition}\end{equation}

The above expression vanishes for ordinary plane waves since $\omega>|k|$
and the detector's speed must be less than unity (=speed of light).
Nevertheless, it can survive for mode functions with Bessel (or Macdonald)
functions, for which $\omega$ can be smaller than $|k|$ locally,
as we will see in the following sections.

However, calculations with Bessel functions in a four dimensional
space is complicated and not easy to understand what is happening
intuitively. Therefore in this section we artificially assume the
mass $m$ is imaginary, i.e., $m^{2}<0$ so that $\omega<|k|$. Although
this is somewhat unrealistic, it can demonstrate how a detector is
excited by negative Killing-energy.

When $\omega<|k|$ the argument of the $\delta$ function in (\ref{transition})
can be non-zero for a large $k$. This means the detector is excited
by the emission of negative Killing energy since the term in (\ref{eq:creation})
is the result of creation operators.

\bigskip{}

In contrast, such negative Killing energy emission does not occur
when we perform the same calculation in the detector's rest frame.
Let us introduce a new frame $\Sigma'$ which is moving with a velocity
$(u_{t},u_{x})=(\zeta_{0t},\zeta_{0x})$ relative to the original
frame (let's call the original frame $\Sigma$); the coordinates in
the frame $\Sigma'$ becomes \begin{equation}
t'=\zeta_{0t}t-\zeta_{0x}x\,,\;\; x'=\zeta_{0t}x-\zeta_{0x}t\end{equation}
 The trajectory of the detector is $(T'(\tau),X'(\tau))=(t',0)$.
When we do the same calculation as above, (\ref{transition}) becomes
\begin{equation}
\int_{-\infty}^{\infty}d\tau\, e^{i\Delta E\tau}\left\langle 1_{k}\right|\phi(T'(\tau),X'(\tau))\left|0\right\rangle =\delta(\Delta E+\omega')\,,\end{equation}
 where $\omega'=|\zeta_{0t}\omega+\zeta_{0x}k|$ is the wave frequency
of the creation operator in $\Sigma'$. The above expression is always
zero since $\omega'>0$.

This discrepancy occurs because the the energy-momentum tensor is
a flux density and its sign depends on the flow direction across the
surface. This situation is illustrated in Figure 1. Two dashed lines
indicate the constant time surface in $\Sigma$ and $\Sigma'$ respectively,
and the hollow arrow is the phase speed of the negative Killing energy
mode in $\Sigma$. The energy-momentum carried by this wave crosses
the constant time surfaces of $\Sigma$ and $\Sigma'$ from the opposite
side, and the flux has opposite sign correspondingly.

\begin{figure}
\begin{centering}
\includegraphics{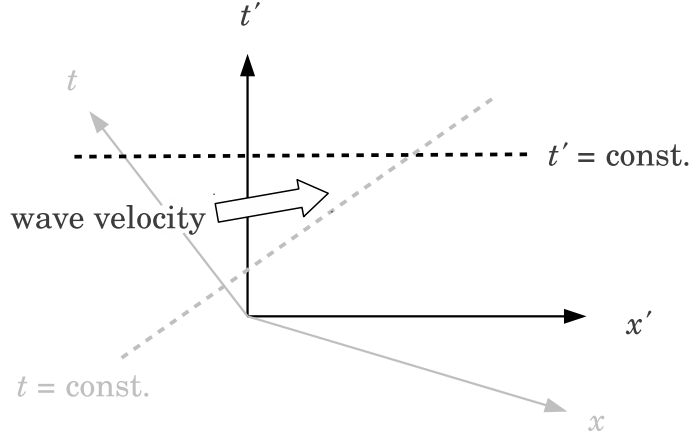} 
\par\end{centering}

\caption{The wave direction crossing the surfaces of constant time. Dashed
lines are the time constant surfaces of $\Sigma$ (constant $t$)
and $\Sigma'$ (constant $t'$) respectively. In $\Sigma'$ the wave
crosses the surface in the increasing $t'$ direction, which is decreasing
$t$ in $\Sigma$. }

\end{figure}

From the above consideration, we understand the negative Killing energy
in $\Sigma$ becomes positive with the same absolute value in $\Sigma'$.
Therefore, if we wish to express the Killing energy in $\Sigma'$
with the parameters defined on $\Sigma$, straightforward Lorentz
transform gives wrong sign. We can obtain the correct answer by replacing
$\zeta_{0t}\omega+\zeta_{0x}k$ with $|\zeta_{0t}\omega+\zeta_{0x}k|$.

\section{Rotating Detector}

In this section we examine the response of a rotating detector. Without
loss of generality we can express the detector's orbit using the proper
time $\tau$ as \begin{equation}
(T(\tau),R(\tau),\Theta(\tau),Z(\tau))=(\gamma\tau,r_{0},\gamma\Omega\tau,0)\end{equation}
 in the cylindrical coordinates $(t,r,\theta,z)$; the radial distance
$r_{0}$ and the angular velocity $\Omega$ are constants, and $\gamma=1/\sqrt{1-\Omega^{2}r_{0}^{2}}$.
The detector cannot move faster than the speed of light, thus $1>\Omega r_{0}$.

The massless Klein-Goldon equation in the cylindrical coordinates
may be written as\begin{equation}
\frac{\partial^{2}}{\partial t^{2}}\phi-\left(\frac{1}{r^{2}}\frac{\partial}{\partial r}r\frac{\partial}{\partial r}+\frac{1}{r^{2}}\frac{\partial^{2}}{\partial\theta^{2}}+\frac{\partial^{2}}{\partial z^{2}}\right)\phi=0\,.\end{equation}
 The mode functions to the above wave equation is \begin{equation}
\psi_{hmk}^{\pm}=\frac{J_{m}(hr)}{\sqrt{8\omega\pi^{2}}}\,\exp-i(\pm\omega-m\theta-kz)\,,\label{eq:bessel}\end{equation}
 where $J_{m}$ is a Bessel function of order $m$ and $\omega=|h|>0$.

The unit vector in the Killing flow direction is expressed as $(\zeta_{t},\zeta_{r},\zeta_{\theta},\zeta_{z})=(\gamma,0,\gamma\Omega r_{0,}0)$.
Then the Killing energy of the mode function across the surface of
constant $t$ (let us call this surface $S$) becomes \begin{equation}
E_{S}=\zeta_{t}T_{t}^{t}-\zeta_{\theta}T_{t}^{\theta}=\gamma(\omega+m\Omega)J_{m}(hr_{0})^{2}\,,\end{equation}
 at the detector's orbit. This can be negative for large negative
$m\Omega$, which may seem peculiar because it means the local frequency
is smaller than the wave number. Suppose we introduce WKB-like approximation
in a small region around $(r,\theta)\sim(r_{0},\theta_{0})$ as \begin{equation}
\psi_{p}^{+}\propto\exp-i(\omega t-k_{r}(r_{0})(r-r_{0})-k_{\theta}r_{0}(\theta-\theta_{0})-k_{z}z)\,,\end{equation}
 The above expression is not quantitative approximation, but just
a rough sketch to illustrate what is happening. The negative $E_{S}$
means $\omega(r_{0})<|k_{\theta}|$ which is not true for ordinary
planar waves since $\omega=\sqrt{k_{r}^{2}+k_{\theta}^{2}+k_{z}^{2}}$.
In this case, however, it can be true because the Bessel function
$J_{m}(hr)$ exponentially damps in $hr\ll1$ for large $m$. Then
$k_{r}(r_{0})$ in the above approximation becomes imaginary to make
$\omega$ smaller than $|k_{\theta}|$. This situation is well mimicked
by the imaginary mass we introduced in the previous section, and the
details we examined there are also valid here.

On the other hand, the Killing energy across the surface normal to
the Killing flow (denoted by $S'$) is \begin{equation}
E_{S'}=\zeta_{t}T_{t}^{t}\zeta^{t}-\zeta_{\theta}T_{t}^{\theta}\zeta^{t}-\zeta_{t}T_{\theta}^{t}\zeta^{\theta}+\zeta_{\theta}T_{\theta}^{\theta}\zeta^{\theta}=\gamma^{2}\omega^{-1}(\omega+m\Omega)^{2}J_{m}(hr_{0})^{2}\,,\end{equation}
 which is always positive. Note that the above Killing energy density
is the one per unit volume. However, energy-momentum of a wave as
a physical entity should be a density per wave length because the
wave length changes due to the Lorentz transform. The Killing energy
density per wave length can be obtained by multiplying the above expression
by a factor $\omega\gamma^{-1}/(\omega+m\Omega)$, which yields the
energy density $\gamma|\omega+m\Omega|J_{m}^{2}$ in consistent with
the result in the previous section.

\bigskip{}

Now we define the detector in the same way as (\ref{detector}) on
$S$. The coefficient for the transition from the Minkowski state
to the one particle state is evaluated as as \begin{eqnarray}
A(E,1_{hmk};E_{0},0_{M}) & = & ic\left\langle E,1_{hmk}\right|\int_{-\infty}^{\infty}d\tau\, m(\tau)\phi(T(\tau),X_{i}(\tau))\left|E_{0},0\right\rangle \nonumber \\
\, & = & ic\left\langle E\right|m(0)\left|E_{0}\right\rangle \int_{-\infty}^{\infty}d\tau\, e^{i\Delta E\tau}\left\langle 1_{k_{hmk}}\right|\phi(T(\tau),X_{i}(\tau))\left|0\right\rangle \,,\label{eq:defA-2}\end{eqnarray}
 where $\Delta E=E-E_{0}$ and $X_{i}(\tau)=(R(\tau),\Theta(\tau),Z(\tau))$
denotes the detector's spatial position.

We expand the field $\phi$ with the mode functions in (\ref{eq:bessel})
as \begin{equation}
\phi=\sum_{hmk}(a_{hmk}^{+}\psi_{hmk}^{+}+a_{hmk}^{-}\psi_{hmk}^{-})\,.\end{equation}
 Mode expansion of the above expression is based on the Hamiltonian
on $S$, therefore, the flux of Killing energy calculated with this
expansion is the one across the surface $S$ as in the previous section.

The terms with annihilation operators vanish for the transition of
$\left|0\right\rangle \rightarrow\left|1_{hmk}\right\rangle $, thus
we have \begin{align}
\int_{-\infty}^{\infty}d\tau\, e^{i\Delta E\tau} & \left\langle 1_{hmk}\right|\phi(T(\tau),X_{i}(\tau)\left|0\right\rangle \nonumber \\
= & \frac{J_{m}(hr_{0})}{\sqrt{8\omega\pi^{2}}}\,\int_{-\infty}^{\infty}d\tau\, e^{i\Delta E\tau}\,\exp(i\gamma(\omega+m\Omega)\tau)\label{eq:delta}\\
= & \frac{J_{m}(hr_{0})}{\sqrt{8\omega\pi^{2}}}\delta(\Delta E+\gamma(\omega+m\Omega))\,.\nonumber \end{align}
 This means the excitation of the detector takes place when $\omega+m\Omega$
is negative; the rotating detector observes particles due to the emission
of negative Killing energy. The amplitude of the above coefficient
is small because the Bessel function becomes exponentially small within
the static limit when $\omega<|m\Omega|$.

\bigskip{}

This excitation by negative Killing energy does not occur when we
choose the surface $S'$ to define the detector as we discussed in
the previous section. Let us introduce new coordinates $(s,\varphi,r,z)$
as \begin{equation}
t=s-r^{2}\Omega\varphi\,,~\theta=\varphi+\Omega s\,.\label{eq:varphi}\end{equation}
 with $r$ and $z$ unchanged. The surface $S'$ is specified by a
constant $s$, which is normal to the $s$ axis.

The mode functions then become \begin{equation}
\psi_{hm'k}^{\prime\pm}=\frac{1}{\sqrt{8\omega\pi^{2}}}\, e^{\mp\omega's}J_{m}(hr)\, e^{im'\varphi}e^{ikz}\,.\end{equation}
 where $\omega'=|\omega-m\Omega|>0$. The filed can be expanded with
these mode functions as

\begin{equation}
\phi=\sum_{hmk}(a_{hm'k}^{\prime+}\psi_{hm'k}^{\prime+}+a_{hm'k}^{\prime-}\psi_{hm'k}^{\prime-})\,.\end{equation}

Since $\varphi$ is constant along the orbit, the same calculation
as in (\ref{eq:delta}) yields \begin{align}
\int_{-\infty}^{\infty}d\tau\, e^{i\Delta E\tau} & \left\langle 1_{hmk}\right|\phi(T(\tau),X_{i}(\tau)\left|0\right\rangle \nonumber \\
= & \frac{J_{m}(hr_{0})}{\sqrt{8\omega\pi^{2}}}\delta(\Delta E+\gamma\omega')\,\end{align}
 The above expression vanishes since the argument of the $\delta$
function is always positive, which means no excitation of the detector.

\section{Accelerating Detector with Drift}

Another class of vacua was investigated in Paper 1. The Killing flow
$K^{\mu}$ to define it is expressed in rectangular coordinates $(t,x,y,z)$
as\begin{equation}
(K^{t},K^{x},K^{y},K^{z})\propto(\Gamma x,\Gamma t,1,0)\,,\label{eq:killing-1}\end{equation}
 which is a flow accelerating in the $xt$ plane superposed with a
constant drift in the $y$ direction. This flow becomes spacelike
when $\Gamma\xi<1$ and therefore, the surface of $\Gamma\xi=1$ is
the static limit. Readers are refereed to Paper 1 for more details;
the above expression is identical to the equation (37) in Paper 1
with $\Gamma=(a^{2}-\omega^{2})/\omega$.

It is reported in Paper 1 that the detector excitation due to the
negative Killing energy takes place here in the same way as for the
rotating detector. We will see that the excitation can be avoided
again when we design the detector appropriately as in the previous
section.

A detector's orbit along the Killing flow (\ref{eq:killing-1}) is
expressed as\begin{align}
T & =\xi_{0}\sinh(\zeta_{0\eta}\tau/\xi_{0})\,,\nonumber \\
X & =\xi_{0}\cosh(\zeta_{0\eta}\tau/\xi_{0})\,,\\
Y & =\zeta_{0y}\tau\,,\; Z=0\,.\nonumber \end{align}
 The parameters $\zeta_{0\eta}$ and $\zeta_{0y}$ are constants corresponds
to the detectors four velocity: $\zeta_{0\eta}=g^{-1}\Gamma\xi_{0}$
and $\zeta_{0y}=g^{-1}$ with $g=\sqrt{\Gamma^{2}\xi_{0}^{2}-1}$.
We can calculate the transition amplitude of the process $\left|E_{0},0\right\rangle \rightarrow\left|E,1_{k_{i}}\right\rangle $
in the same way as the previous section as\begin{eqnarray}
A(E,1_{k_{i}};E_{0},0) & = & ic\left\langle E,1_{k_{i}}\right|\int_{-\infty}^{\infty}d\tau\,\mu(\tau)\phi(T(\tau),X_{i}(\tau))\left|E_{0},0\right\rangle \nonumber \\
\, & = & ic\left\langle E\right|m(0)\left|E_{0}\right\rangle \int_{-\infty}^{\infty}d\tau\, e^{i\Delta E\tau}\left\langle 1_{k_{i}}\right|{\phi}(T,X_{i})\left|0\right\rangle \,,\label{eq:defA}\end{eqnarray}
 where $\Delta E=E-E_{0}$ and $(t,x,y,z)=(T(\tau),X(\tau),Y(\tau),Z(\tau))=(T(\tau),X_{i}(\tau))$
is the detector's trajectory; $\left|1_{k_{i}}\right\rangle $ is
the state one particle with mode $k_{i}$. The field $\phi$ is expanded
as\begin{equation}
\phi=\left(a_{k_{i}}^{+}\, e^{-i\omega t}+a_{k_{i}}^{-}\, e^{i\omega t}\right)e^{k_{i}x^{i}}\,.\end{equation}
 Again this expression means that the Killing energy is the one across
the surface of constant $t$ as in the previous section. With the
above expansion we obtain\begin{align}
\int_{-\infty}^{\infty}d\tau & e^{\Delta E\tau}\left\langle 1_{k}\right|\phi(T(\tau),Z(\tau),Y(\tau),Z(\tau))\left|0_{M}\right\rangle \nonumber \\
= & \frac{1}{\sqrt{2\pi\omega}}\int_{-\infty}^{\infty}d\tau\, e^{i\Delta E\tau}\exp i(\omega T(\tau)+k_{i}X^{i}(\tau))\end{align}

To simplify the detector's orbit we introduce the Rindler coordinates
for the $tx$ plane as\begin{equation}
t=\xi\sinh\kappa\eta,\; x=\xi\cosh\kappa\eta\;,\label{eq:rindler-1}\end{equation}
 where $\kappa$ is an arbitrary constant to make the arguments of
hyperbolic functions dimensionless so that $\eta$ has the unit of
length. The detector's orbit is expressed as $(\eta,\xi,y,z)=(\kappa^{-1}\xi_{0}^{-1}\zeta_{0\eta}\tau,\xi_{0},\zeta_{0y}\tau,0)$
with these coordinates. The mode functions in this coordinate system
is expressed as \begin{equation}
\psi_{p}^{\pm}=\frac{\sqrt{\sinh\sigma\kappa^{-1}}}{2\pi^{2}}\, e^{\mp\sigma\eta}K_{ip}(h\xi)\exp i(k_{y}y+k_{z}z)\,.\end{equation}
 where $K_{ip}$ is the Macdonald function (modified Bessel function)
with the imaginary order and $\sigma=|p|,\, h=\sqrt{k_{x}^{2}+k_{y}^{2}}$.

The Minkowski modes can be expanded by the above mode functions as\begin{equation}
\frac{1}{\sqrt{2\pi\omega}}\exp-i(\pm\omega t-k_{i}x^{i})=\sum_{p}[\alpha(p,k_{i})\psi_{p}^{\pm}+\beta(p,k_{i})\psi_{p}^{\mp}]\,.\end{equation}
 where $\alpha$ and $\beta$ are the Bogolubov coefficients conventionally
used to calculate the Unruh effect. With the above expansion we obtain
\begin{align}
\int_{-\infty}^{\infty}d\tau & \, e^{i\Delta E\tau}\exp i(\omega T(\tau)+k_{i}X^{i}(\tau))\nonumber \\
\, & =\int_{-\infty}^{\infty}d\tau\int_{-\infty}^{\infty}dp\, e^{i\Delta E\tau}e^{ik_{y}\zeta_{0y}\tau}\left[\alpha(p,k_{i})\,\exp\left(\frac{i\sigma\zeta_{0\eta}}{\kappa\xi_{0}}\tau\right)\right.\nonumber \\
 & ~~~~~~~~~~~~~~~~~~~~~~~~~~~\left.+\beta(p,k_{i})\,\exp\left(-\frac{i\sigma\zeta_{0\eta}}{\kappa\xi_{0}}\tau\right)\right]K_{ip}(h\xi_{0})\,\nonumber \\
\, & =\int_{-\infty}^{\infty}dp\,[\alpha(p,k_{x})\,\delta(\Delta E+g(\kappa^{-1}\Gamma\sigma+k_{y}))\nonumber \\
 & ~~~~~~~~~~~~~+\beta(p,k_{x})\,\delta(\Delta E-g(\kappa^{-1}\Gamma\sigma-k_{y}))]K_{ip}(h\xi_{0})\,\,.\end{align}
 The terms with Bogolubov coefficients $\beta$ is the result of annihilation
operators, which means the absorption of quanta excites the detector
as in the usual Unruh effect. The excitation by negative Killing energy
is expressed by the terms with coefficients $\alpha$ as in the previous
section.

These terms with $\alpha$ again vanish when we choose the surface
normal to the Killing flow of the detector's orbit. Actual calculation
is similar to the one in the previous section. Or, we can obtain the
same result simply by replacing $\kappa^{-1}\Gamma\sigma\pm k_{y}$
with $|\kappa^{-1}\Gamma\sigma\pm k_{y}|$ following the prescription
in Section II. The result shows coefficients with $\alpha$ vanish
but those with $\beta$ survive. This means the detector responds
not by the negative Killing energy emission, but by the absorption
of positive Killing energy only, as in the usual Unruh effect. Further
calculation leads to the particle distribution of Doppler shifted
Planckian distribution as expected.

\section{Concluding Remarks}

In the present paper we have investigated the vacuum observed by a
circularly rotating Unruh-DeWitt detector. The response of a detector
depends on the choice of the surface (three volume) for the Hamiltonian
to define it. Consequently detectors defined on different surfaces
may perceive different state of particles. The reason for the particle
detection reported in the past literature is due to the choice of
the surface with constant Minkowski time. A detector will not observe
particles when we define it on a surface normal to the detector's
orbit.

It has been puzzling that a rotating detector observed particles in
a Minkowski vacuum because a global analysis shows the rotating vacuum
is identical to the Minkowski vacuum. Korsbakken and Leinaas \cite{Korsbakken2004}
clarified the reason for this discrepancy. They found the detector
responds to the negative Killing energy wave; the ground state detector
can get excited by the emission of negative Killing energy mode. In
the present paper it was shown that their result is due to the choice
of surface to define the detector; their choice was the surface of
constant Minkowski time. Here in the present paper we introduced a
detector defined on a surface normal to the detector's orbit. It was
found such a detector does not perceive negative Killing energy, and
thus particles are not detected. A similar situation was also found
for an accelerating detector with perpendicular drift.

Using a hypothetical negative mass, we demonstrated how and why negative
Killing energy occurs. When the phase speed of some waves is slower
than the detector, such waves crosses the surface of the constant
time from the {}``flip side'' of the surface. Consequently, the
energy-momentum flux has opposite sign, since the sign of flux is
determined by the direction of surface to cross. The detector sees
negative Killing energy for those waves, and can be excited by the
absorption of negative Killing energy.

A remark should be made that the definition of the Hamiltonian in
the present paper is not rigorous in a sense. Precisely speaking,
a surface of a Hamiltonian for field quantization must be all spacelike,
however, the surface we introduced here becomes timelike beyond the
static limit. There are attempt to generalize the field quantization
to accommodate such partially timelike surface (see \cite{Oeckl2006}
and references therein), however, it is out of scope of the present
paper to discuss it. We simply assume its validity here. Also there
is a subtle point at defining the constant time surface with the coordinates
(\ref{eq:varphi}), which has discontinuity at $\theta=2\pi$. We
will leave detailed examination on this point for future work.

\thanks{ The author is grateful to MANZANA for productive research environment. }

\bibliographystyle{apsrev4-1} \bibliographystyle{apsrev4-1} \bibliographystyle{apsrev4-1}
\bibliography{Unruh}

\end{document}